\documentclass[conference]{IEEEtran}
\IEEEoverridecommandlockouts
\usepackage{cite}
\usepackage{amsmath,amssymb,amsfonts}
\usepackage{algorithmic}
\usepackage{graphicx}
\usepackage{textcomp}
\usepackage{xcolor}
\usepackage{lipsum}
\def\BibTeX{{\rm B\kern-.05em{\sc i\kern-.025em b}\kern-.08em
  T\kern-.1667em\lower.7ex\hbox{E}\kern-.125emX}}
\begin{document}

\renewenvironment{quote}%
  {\vspace{5px}\list{}{\leftmargin=0.8em}\item[]}%
  {\endlist\vspace{5px}}

\title{Expanding AI Awareness Through Everyday Interactions with AI: A Reflective Journal Study\\

\thanks{Funding: U.S. NSF Awards 2319137, 1954556, USDA/NIFA Award 2021-67021-35329, and GMU MARIE Program.}
}

\author{
\IEEEauthorblockN{Ashish Hingle}
\IEEEauthorblockA{\textit{Information Sciences and Technology} \\
\textit{George Mason University}\\
Fairfax, USA \\
ahingle2@gmu.edu}
\and
\IEEEauthorblockN{Aditya Johri}
\IEEEauthorblockA{\textit{Information Sciences and Technology} \\
\textit{George Mason University}\\
Fairfax, USA \\
johri@gmu.edu}
}

\maketitle

\begin{abstract}
This \textbf{research} paper presents findings from a reflective journal study of undergraduate students' everyday interactions with artificial intelligence (AI). As the application of AI continues to expand, students in technology programs are poised to be both producers and users of the technologies. They are also positioned to engage with AI applications within and outside the classroom. While focusing on the curriculum when examining students' AI knowledge is common, extending this connection to students' everyday interactions with AI provides a more complete picture of their learning. In this paper, we explore student's awareness and engagement with AI in the context of school and their daily lives. Over six weeks, 22 undergraduate students participated in a reflective journal study and submitted a weekly journal entry about their interactions with AI. The participants were recruited from a technology and society course that focuses on the implications of technology on people, communities, and processes. In their weekly journal entries, participants reflected on interactions with AI on campus (coursework, advertises campus events, or seminars) and beyond (social media, news, or conversations with friends and family). The journal prompts were designed to help them think through what they had read, watched, or been told and reflect on the development of their own perspectives, knowledge, and literacy on the topic. Overall, students described nine categories of interactions: coursework, news and current events, using software and applications, university events, social media related to their work, personal discussions with friends and family, interacting with content, and gaming. Students reported that completing the diaries allowed them time for reflection and made them more aware of the presence of AI in their daily lives and of its potential benefits and drawbacks. This research contributes to the ongoing work on AI awareness and literacy by bringing in perspectives from beyond a formal educational context. 
\end{abstract}

\begin{IEEEkeywords}
AI literacy, journal study, reflection-based assessment
\end{IEEEkeywords}

\section{Introduction}
Artificial intelligence (AI) driven applications are becoming increasingly common across domains with significant influence over societal functions, including dynamic pricing of consumer goods, search results, and how dependent future learners are on tools \cite{buchh2022dynamicprice, noble2018algorithmsopression, zhao2024chatgpt}. Some examples of our everyday interactions with AI include "recommender systems, customer service chatbots, search engine algorithms, smart assistants, digital self-tracking services, etc." \cite[p.~127]{lomborg2022everyday}. While they may take many forms, the growing prevalence of AI systems highlights the importance of developing an AI-aware and literate population, especially using a "high-level understanding" as a baseline for the general population \cite{long2021alliteracyexhibit}. Developing AI awareness is especially important for current students, given it will likely affect their day-to-day interactions. Leveraging students' personal experiences as educational tools can effectively enhance their understanding of AI. While formal AI education programs are increasingly being developed to reach a wider audience, exploring and incorporating informal interactions with AI technologies is important to provide a more comprehensive understanding \cite{ng2021aisocialnetwork, long2021alliteracyexhibit}.

This study explores students' self-reflections on their interactions with AI by bridging their in-class experiences through a technology ethics course with their informal interactions with everyday AI. We implemented a weekly reflective journal for undergraduate technology students, asking them to document and reflect on their interactions with AI over six weeks. The theoretical framing for this work is based on prior work on computing and technology students learning through formal and informal activities \cite{mccartney2016selfdirected, boustedt2011formalinformal} and the use of journals for student self-reflection on AI interactions \cite{koenig2020literacyjournals}. The research questions guiding this study, which were developed through reviewing categorizations of AI and their use \cite{long2020ailiteracy, kasinidou2023aicourse, hornberger2023instrument}, were:
\begin{enumerate}
  \item RQ1: Which everyday interactions with AI are salient for students in the classroom and outside?
  \item RQ2: How did students perceive the process of reflecting on AI interactions?
\end{enumerate}

\section{Relevant Prior Work}

\subsection{AI Awareness, Literacy, and Learning}

The field of education has been profoundly impacted by everyday AI, particularly through its integration into routine and often subtle daily interactions for learners and educators alike \cite{lomborg2022everyday, johri-ambivalence-2023}. As researchers work to define and integrate AI learning, considerable research has been dedicated to exploring AI concepts, methods of implementation, and assessment approaches. Several reviews have addressed AI literacy (or similar constructs questioning what people need to know about AI) in education contexts and have highlighted the state of the teaching and learning landscape over the last few years \cite{long2020ailiteracy, ng2021conceptualizing, casal2023ai, almatrafi2024ailiteracy}. Quantitative AI literacy instruments are also increasingly common across these constructs. Despite a variety of framings, the literature consistently differentiates between technical and non-technical AI skills, emphasizing the importance of understanding how AI works and addressing the societal implications of AI systems \cite{carolus2023mails, hornberger2023instrument}.

While AI learning initiatives increase, there is room for further discussion connecting students' learning in the classroom to activities beyond. Students' understanding of what AI is and how it affects them may be of concern \cite{hingle2023conceptions}. Researchers have attempted to make these connections apparent through various interventions. Ali et al. developed a card game that describes uses of AI that could be familiar to learners and focuses on having students actively think about the harms caused by AI systems and ways to mitigate them \cite{ali2023aiaudit}. Kasinidou et al. describe a course implemented at the Open University of Cyprus that included eight units addressing various AI applications, including computer vision, natural language processing, ethics, and living with AI \cite{kasinidou2023aicourse}. These initiatives integrate learning experiences, and more research must explore how students engage and learn from AI in the wild. 

Informal interactions with AI are gaining traction as an effective means to educate the public about AI concepts outside traditional classroom settings. These activities can include efforts to connect to everyday practices and settings to make AI knowledge accessible and engaging \cite{hingle2024framework}. A study by Long et al. highlights the importance of incorporating creative and embodied interactions in informal learning spaces, such as museum exhibits, to foster interest in AI and computing-related topics \cite{long2021alliteracyexhibit}. The study proposes design principles for informal settings, emphasizing collaboration and creativity in making AI concepts relatable and understandable. By integrating informal learning activities, these programs can reach a broader audience, thus helping to demystify AI and make it more accessible to the general public \cite{long2023explainability}.

\subsection{Student Reflection and Journals for Informal Learning}

Student reflection activities are defined broadly across the literature as creating space to think through questions, ponder alternatives, and synthesize knowledge meaningfully for the learner. Clarà highlights a central component from Dewey's earlier work (1933) of reflection to include observations and inferences: "Observation operates by means of facts, the objective events of a situation, what Dewey sometimes calls conditions or data, whereas inference is a jump beyond these observed events, it is an act of imagination, a supposition or hypothesis, what Dewey sometimes calls suggestion or idea." \cite[p.~265]{clara2015reflection}. For students, the reflection can involve strategies to help them think like a professional in their field by bridging their experiences with how they think of the profession and learning about it \cite{tanner2012metacognition}. Navigating what was experienced, grappling with the possibilities, mulling through alternative paths, and building informed opinions is a significant portion of the reflective experience.

Researchers have explored the effectiveness of reflection by analyzing the key components that constitute high-quality reflective practices. Lin et al. describe three processes in reflective thinking and designing technology-based environments that support reflection: "reflection 1) involves social interactions; 2) is an active, intentional, and purposeful process of exploration, discovery, and learning; and 3) involves understanding one's process of learning" \cite{lin1999designing}. These definitions can be mapped to Chickering and Gamson's seminal text on \textit{The Seven Principles for Good Practice in Undergraduate Education}. Reflection can be an active learning activity (Principle 3), provide students the temporal opportunity to spend time on the task (Principle 5), and engage the diverse talents of the learner by allowing for a variety of media in the reflection (Principle 7). By constructing reflection through this lens and focusing on an appropriate level of reflection, i.e., simulated, descriptive, dialogic, or critical \cite{strampel2007using}, learners can be given structure even in an exercise that can seem fluid.

Reflective activities can take various forms: in class, outside, or based on assignments, for example. Reflective journals are one such data collection technique that allows researchers to collect data from participants over time as they happen. They are similar in structure to diary studies. Carter and Mankoff describe feedback studies, where the participants respond to predefined questions, and elicitation studies, where participants capture thoughts through writing and media, which can be used in the post-study follow-up \cite{carter2005roleofdiary}. The data is collected while not directly involved with the participants to minimize the effects of being observed in the way an interview may influence engagement. Journal studies are a significant time and effort activity for participants, and it is encouraged to provide compensation for the engagement.

Some researchers have used reflective activities specifically within the context of algorithmic literacy and technology awareness. Koenig describes using reflective student media journals to have communication students reflect on their engagements with algorithms \cite{koenig2020literacyjournals}. Students were prompted to explore their algorithmic influences and literacy through everyday activities. Koenig highlights students moving toward more rhetorical and critical discussions through their media journals by exploring basic, rhetorical, and critical awareness through the journals. "As my results suggest, because the students were asked to write journal entries on this topic, they were required to not just consider their actions, but turn those actions into words. Because they were asked to examine those patterns of use and their own role in creating those patterns, students gained critical and rhetorical understanding of their engagements" (pg. 11).

\section{Method}
In this study, participants submitted their thoughts, perceptions, and opinions on their interactions with AI through a reflective journal. Participants completed one reflective journal entry weekly for six consecutive weeks, starting in October 2023 and ending in November 2023. The journals were collected as part of a more extensive study on students' everyday AI interactions. The primary methodology used a feedback approach to elicit participants' descriptions of their interactions with AI. However, due to the open-ended nature of the prompts, which encouraged thorough and expansive responses, an elicitation component was also incorporated \cite{carter2005roleofdiary}.

Participants were provided with weekly prompts designed to center the discussion around interactions broadly with AI. Participants were encouraged to describe the formal (structured, sequential, repressive, and compulsory learning that is teacher-led and assessed), non-formal (occurs outside of the school and is also structured but is non-sequential, non-repressive, and voluntary learning that teachers may guide), and informal (unstructured, spontaneous, and usually learner-led, allowing the learner to decide when and how it occurs) interactions \cite{eshach2007bridging}. The prompts given to students were inspired by broad categorizations of AI explored through recent AI literacy systematic literature reviews \cite{almatrafi2024ailiteracy, long2020ailiteracy, ng2021conceptualizing}. Specifically, students were instructed to submit a journal entry in response to the following prompt:

\begin{enumerate}
    \item 1. In the last week, document where and about what you have noticed mention about artificial intelligence (AI):
    \begin{enumerate}
        \item at your campus (advertised events, courses you are taking, campus news, or guest speakers) or
        \item outside of your campus (your social media, local or international news, conversations with family)?
    \end{enumerate}
    \item Was the way AI was described leave you with a positive, negative, or neutral impression? Why or why not?
\end{enumerate}

We chose to undertake this research as a qualitative study with a comprehensive approach to capture a range of student responses. This decision reflects the novelty of investigating informal learning activities among technology and computing students, particularly in light of recent advancements in AI, such as LLMs, that have significantly increased the accessibility of consumer-facing AI. Reviewing the literature, we found few studies exploring how students reflect on their interactions with AI in and outside the classroom. However, our methodology follows guidelines for conducting research from similar approaches \cite{mccartney2016selfdirected, boustedt2011formalinformal}.

\subsection{The Course}
Students were recruited to participate from a technology ethics course hosted by a College of Engineering and Computing at a large, public university in the Southeastern United States. The course covers the social implications of new technology. Students learn about how technology affects systems on an individual, national, and global scale. The course is structured into several modules, consistently focusing throughout the 16-week semester on societal implications, ethics, and values in the design and experience of an interconnected world. The course brings students from various disciplines together and addresses \textit{ABET Accreditation Outcome 3} on Student Outcomes.

\subsection{Participants}
Twenty-two (22) students were recruited from the course to participate in this study in October 2023. Participants ranged from 19 to 32 years old and completed between 40 and 121 university credits. Eleven (11) participants self-described their gender as female, and eleven (11) participants self-described as male. The majority of the students were full-time, working towards a BS in Information Technology with a variety of specializations.

Participants were provided a \$48 gift card for completing the six journal entries. No course credit was offered for participating, and deciding not to participate did not impact course outcomes.

\begin{table*}[]
\centering
\caption{Number of Interaction Items across All Journal Entries}
\setlength{\tabcolsep}{10pt}
\renewcommand{\arraystretch}{1.5}
\begin{tabular}{|p{1.1in}|p{1.65in}|p{2.6in}|p{0.4in}|}
\hline
\textbf{Interaction} & \textbf{Description} & \textbf{Example from Journals} & \textbf{Journal Freq.} \\
\hline
\textbf{Coursework}\newline (Formal Learning) & Discourse and discussions from Technical and non-technical course readings, activities, and lectures.& "In the course, the depth of the conversation around deep fakes was both illuminating and alarming." Participant 18, Week 3. & 49 \\
\hline
\textbf{News and Current Events}\newline (Informal Learning) & Describing reading or watching a news piece. Some participants highlighted doing so for a course, while others did not make this distinction.& "I saw a news article about an executive order from the White House that is forthcoming. I think it is great for regulations to be included in the overall discussion of AI." Participant 11, Week 2 & 47 \\
\hline
\textbf{Using Software and\newline Applications}\newline (Informal Learning) & Experiences using software and applications, including ChatGPT, Spotify, Amazon's Customer Service Chatbot, and algorithmic recommenders, etc.& "I have been using a few AI tools like ChatGPT and Midjourney to experiment with for my work." Participant 22, Week 5 & 22 \\
\hline
\textbf{University Events}\newline (Formal Learning) & Interactions at an event organized by the university or college, such as a research talk, research series, fireside chat, workshop, hackathon, etc.& "This week, I heard about a series of workshops through my faculty that will be held at Harvard on building AI models for vision tasks specifically. We were invited to participate and learn, and I probably will go." Participant 21, Week 4 & 14 \\
\hline
\textbf{Through Social Media}\newline (Informal Learning) & Discourse of AI interactions on social media platforms, including Facebook, Instagram, TikTok, Snapchat, etc.& "On social media I see a lot of advertisements from companies like Adobe showcasing their AI generative tools in their products. Even though I don't use Photoshop, nor do I follow that content, because of its AI generative tools it was brought to my attention as an AD suggestion." Participant 6, Week 2 & 13 \\
\hline
\textbf{At Current Job or about Future Employment}\newline (Informal Learning) & Highlighting discourse at work, current affairs in the participant's employment, or with coworkers about AI.& I was very interested in [restaurant automation through AI], because I used to work at Starbucks, and I know that Starbucks is also planning to utilize AI for their machines. The new espresso machines that you see at some stores are called the Mastrena 2, which provides telemetry data to Starbucks' support center." Participant 6, Week 3 & 10 \\
\hline
\textbf{Discussions with Friends \& Family}\newline (Informal Learning) & Discourse and Discussions with friends and family. Can be instructive or just conversational. & "During conversations with family members, the subject of AI's effects on the labor market came up, with some voicing worries about future job displacement as a result of automation." Participant 4, Week 1 & 9 \\
\hline
\textbf{Interacting with Content}\newline (Informal Learning) & Discourse related to content (media created for an audience to consume), including YouTube Videos, Content Creators, etc.& "But I have learned that people on Threads are also talking about [the new Spotify AI DJ feature] as well, specifically a thread started by MKBHD (a tech reviewer on YouTube) and he has stated that after testing it works well for him." Participant 18, Week 4 & 6 \\
\hline
\textbf{Gaming}\newline (Informal Learning) & Related to video game development and game-play experiences.& "I spent a lot of time playing Mario Kart 8 Deluxe and while I enjoyed my time with it I started disliking the rubber banding from the CPU opponents, especially at the end of the race where several time they appear to blatantly stop and let me through to win, making me wish the AI was more predictable." Participant 3, Week 1 & 3 \\
\hline
\end{tabular}
\end{table*}

\subsection{Analysis}
Twenty-two participants completed and submitted a total of 132 reflective journal entries - each student completed 6 journal entries over the course of 6 weeks. The authors analyzed the student reflections through an emergent analysis method to explore a broad perspective on the discourse students entered in their journals, similar to the analysis framing highlighted in \cite{koenig2020literacyjournals}. The authors read through the journal entries and free-coded the journal entries, from which a set of emergent themes and subthemes were identified. Initially, this resulted in 15 themes, and after discussing the codes in each, rereading and identifying collapsible codes, and ensuring saturation was reached, a final set of 9 themes was created (Table I).

\section{Findings and Discussion}

\subsection{Which everyday interactions with AI are salient for students in the classroom and outside? (RQ1)}
In addressing RQ1, we explore the interactions students described in their weekly journal entries. The analysis revealed the emergence of 9 themes. While frequency is one method of understanding these interactions, it is important to recognize the significance of those interactions with fewer instances, as they highlight the ways in which students engage with the subject matter across formal, informal, and non-formal contexts. Table I describes the results across all the journals. We present more details for the themes in the following sections and how students explored the aspects of their learning.

\subsubsection{\textbf{Coursework}}
Participants primarily focused their reflections on the course discussions when describing their interactions throughout the week. This observation could be attributed to several factors. Firstly, the course held a central position within the study design; it was the formal learning component and where the students were recruited from. Secondly, the nature of the prompt likely influenced this inclination, as it encouraged participants to view the course as a potential locus of interaction. This approach was deliberately crafted to ensure that all participants would have something to draw upon during a slow week.

Across the journal entries, the most common reflection was by participants discussing what was raised in the course and reflecting on what they had learned. Participant 18 described discussing the topic of deep fakes in a structured way in the classroom and used the journal to build on both the benefits and the problems of the technology: 

\begin{quote}
\textit{In [the course], the depth of the conversation around deep fakes was both illuminating and alarming. On one hand, the beneficial aspects of deep fake AI are there. The technology can revolutionize the entertainment industry, allowing filmmakers to recreate past performances or generate realistic visual effects without the need for costly CGI. Additionally, in educational contexts, deep fakes could potentially bring historical figures "back to life," providing a dynamic way to engage with history. However, alongside these potential benefits are negatives. The most pressing concern is the misuse of deep fakes in spreading misinformation or fake news, potentially swaying public opinion or causing unwarranted panic.} [Week 3]
\end{quote}

With these types of reflections about the course materials, the timeliness and relevance beyond the classroom of the course discussion became apparent. Deepfakes are a visual form of disinformation that can affect cognitive outcomes, reduce trust in news on social media, and generate uncertainty \cite{vaccari2020deepfakes}. Though not initially dubbed an everyday interaction, the ease of developing and deploying these systems has grown exponentially, and participants were generally concerned about how these might alter social dynamics.

This pattern of AI's societal implications was common across the course learning. Participant 6 reflects on not just learning about the problem but the types of solutions emerging from the discussion:

\begin{quote}
\textit{The [course] group also discussed about ethics that AI brings in art. A lot of artists believe that AI simply steals artwork from people and enables those who use it to claim artwork as their own. I understand and agree slightly with this because artists who work hard developing a skill are affected by this. My classmates told us that there are tools that help artists fight back against AI by poison AI data which is called Nightshade.} [Week 5]
\end{quote}

For the majority of participants, the formal course engagement was vital for raising literacy and knowledge on the topics but it also highlighted how collaborative engagement on the topic is important to learn about these issues.

Other participants also raised course interactions beyond the course they were recruited from. Participant 17 described engaging with the AI discussion for a semester-long research paper:

\begin{quote}
\textit{"In my English class, we are supposed to research information in accordance with our discipline and myself along with a lot of the other classmates have chosen AI to be our topic of choice for our research papers."} [Week 1]
\end{quote}

Though this is expected, ensuring discussions about AI are presented in the context of different fields is fundamental to ensuring students have a holistic picture.

\subsubsection{\textbf{News and Current Events}}
Participants often raised news and current events through relevant articles at the time of data collection (October-November 2023). These came from both formal and informal spaces; sometimes, the articles were referenced as part of their coursework, but in other cases, the news or current events were described in the wild.

For example, Participant 11 writes about coming across discourse on regulations for responsible AI use after becoming primed to the conversation through course discussions:

\begin{quote}
\textit{I saw a news article about an executive order from the white house that is forthcoming. I think it is great for regulations to be included in the overall discussion of AI. We really need new ways of thinking about this and there need to be government regulations that encourage and enforce them. I am neutral on AI, but positive on regulations.} [Week 2]
\end{quote}

The reflection on regulations across AI development is not new but was ever-present across the journals. Most Americans already support oversight and regulation in the ways the technology is used \cite{faverio2023americansviewsAI}. Having students engage with the discussion and arrive at their own consensus is a reflective exercise.

Discussions about the negatives were also tempered with the positive as some students engaged with a positive lens on the uses of AI. Participant 5 described positive news about AI and how it felt out of place in the larger discussion of news and current events:

\begin{quote}
\textit{[AI seems] to have a knack for spotting patterns that might easily slip past the human eye. What struck a chord with me was the sense of hope [the article I read] radiated. It was heartening to see AI framed not just as a flashy piece of tech but as a genuine ally in our uphill battle against climate change.} [Week 3]
\end{quote}

The article linked by the participant focused on predicting climate shifts when facing extreme weather and explored the differences between traditional and machine learning methods, highlighting the ability to work with complex data \cite{deloss2023aiweather}. The participant came across this article on their own and built their understanding of the foundations of what was discussed in the course.

Regarding a news piece about the UN's role in addressing AI, Participant 7 projects the ideas presented through the article on the past, exploring the implications on the impact of people:

\begin{quote}
\textit{I thought it was very interesting because there is always a discussion about AI being used for bad along with the good. I think this is really important. If we don't have these discussions now, the technology will get out of hand and we will have to talk about the issues as they are actively affecting people kind of like if we talked about the impact of Facebook before it grew so much, maybe less people would be hurt by some of their practices?} [Week 1]
\end{quote}

Ensuring participants reach this level of introspection is important in discussing how ethics and social discourse affect learning.

\subsubsection{\textbf{Using Software and Applications}}
Reflections of interactions with software and applications emerged frequently and highlighted participants' navigating learning about AI while using it. Participant 19 highlights their engagements with \textit{Spotify} to recommend new artists and music:

\begin{quote}
\textit{My favorite feature of AI is through Spotify. They have a playlist called a "daylist" that updates roughly 5 times a day to match exactly what kind of music you like to listen to at that time of day mixed with artists that you listen to regularly and similar new music and artists you may enjoy.} [Week 1]
\end{quote}

Though not explicitly highlighted, the recommendations are aligned with musical taste preferences based on previous usage. We found that participants who engaged with recommendation services employed them for various tasks. In the following week's journal, Participant 19 highlights their search for recommendations on \textit{Instagram} when visiting a new place:

\begin{quote}
\textit{My second encounter was via Instagram Reels, I was visiting Salem, Massachusetts over fall break and needed more activities to do but was having difficulty finding some online. To get better access to information on my trip such as, where to eat, cool places to visit, etc. I looked up Salem on Instagram and liked a few reels related to Salem to help my algorithm know that that's what I was looking for. After liking 3-4 posts relating to Salem, my feed was able to learn that this was the content I wanted and provide me with great suggestions.} [Week 2]
\end{quote}

This type of information-seeking behavior is common across social media platforms where these tools serve as community engagement and credibility, though may be standardized, can provide users with quick, less important information easily \cite{marcella2023purposive}. Participants highlighted these interactions frequently across the journal entries.

In contrast to deciding to use an application, several participants highlighted involuntary interactions, often through campus-mandated tools. Online exam proctoring has been a widely discussed topic since the technology has become a staple across coursework over the last decade. Participant 1 reflected on their awareness of the algorithms in the system and highlighted their discontent: 

\begin{quote}
\textit{Last week, I had mentioned that I was helping a professor with [online proctoring software] after realizing there was a problem with the system. I think he agreed to reconsider using it because he said going forward, we'll try to do this exam in class. Personally, I like this more. The impression I had of the [online proctoring software] was overall negative and I would rather not use it if I can avoid it.} [Week 4]
\end{quote}


\subsubsection{\textbf{University Events}}
Universities throughout the US are embracing various approaches to incorporate artificial intelligence into their teaching and learning frameworks. A prevailing trend across institutions involves establishing programs and events designed to expose students to AI concepts during their time as students. Following this trend, the institution where the participating students were recruited has implemented several initiatives, including guest speaker events, fireside chats, and hackathons to foster engagement with AI among its student body. As Participant 17 describes, the levels across discussion can range and can serve as motivation (or potentially barriers) for non-formal learning:

\begin{quote}
\textit{I saw a post from the Statistics department about an AI seminar called "Functional Data Analysis Through the Lens of Deep Neural Networks". I thought the topic was really interesting and shows the type of work that neural networks are possibly able to do. It did look very complicated though so maybe I still have a lot to learn here.} [Week 4]
\end{quote}

While some students found these events useful, others, such as Participant 21, highlight what might come off as performative organization or following buzzwords:
\begin{quote}
\textit{I have heard about AI a lot on campus, but tbh, most of it is not that memorable. AI feels like a buzzword that we hear of often across the campus, maybe because it comes with being a technology or engineering department.} [Week 1]
\end{quote}

Ultimately, university events serve as non-formal events that can introduce and inspire students. Engaging meaningfully is fundamental to ensure students are making the most of the experience.

\subsubsection{\textbf{Through Social Media}}
Participants described social media as a significant source of information for them. As such, several of the interactions highlighted by the students were linked to what showed up on their feeds, highlighting the filter bubbles that can exist on one's social media. Participant 14 describes watching a video on deep fakes on their feed:
\begin{quote}
\textit{The notion that AI could replicate one's identity to fabricate false narratives or malicious activities is deeply unsettling. The video's portrayal of deep fakes and AI manipulation underscored the vulnerabilities inherent in our increasingly digitized world. It emphasized the need for heightened vigilance in discerning between authentic and fabricated content, especially in an era where the spread of misinformation can have profound societal and political consequences.} - [Week 5]
\end{quote}

\subsubsection{\textbf{At Work or about Employment}}
Several students discuss the similarities and differences between discussions about AI in their classes and at work. The workplace is often where students must demonstrate their training and knowledge of concepts while adapting to the needs of the work. Participant 12 reflects on contrasting viewpoints of AI between what they are learning about in the classroom and outside:

\begin{quote}
\textit{In my course, the focus was on the impact of AI in defense. The discussions leaned towards the negative side, as we delved into the ethical and moral dilemmas posed by AI in military applications. This context raised concerns about the potential misuse of AI technology and its consequences, leaving me with a somewhat pessimistic view of AI in this specific context. Conversely, at my workplace, there was excitement and positivity surrounding AI. We are in the process of launching an AI-generated tool to assist us in searching for essential information within our vast internal resources. The prospect of increased efficiency and productivity through AI was met with enthusiasm, as it promises to simplify our workflow and enhance our ability to access and utilize valuable data.} [Week 1]
\end{quote}

Other students were interested in what the future of their own work would be. Participant 22 questions how generative AI specifically would affect the ways they would create in the future:

\begin{quote}
\textit{Through my work as a social media intern, I have been thinking about how AI is used to make content especially thinking about the future of my work. Some of the people at work use ChatGPT to edit, copy, and write short pieces that are later edited, so it is interesting to think about what the future of this work will be.} [Week 1]
\end{quote}

Participant 6 highlights being concerned about what the implications of automated work would look like:
\begin{quote}
\textit{The new [Starbucks] espresso machines that you see at some stores are called the Mastrena 2, which provides telemetry data to Starbucks' support center. This allows the support center to track maintenance data on machines, which could automatically send out parts or even mechanics to service without employees having to call. They can also utilize the data they receive to track the quality of the shots and how many espresso drinks they're making. It left me with a positive impression because I know that it will help utilize efficiency in the stores, although I am a little worried about the consequences that could result.} [Week 3]
\end{quote}

\subsubsection{\textbf{Discussions with Friends and Family}}
Discussions between the participants and their friends and family members emerged across the journal entries. For some, like Participant 13, it was in understanding the practical implications of building AI systems:
\begin{quote}
\textit{I was talking with my father who works as a salesman at BMW about the technology in the cars and how advanced they are getting. He was telling me about how difficult it is to repair these items as well, and that often the service center ends up having to replace the whole unit when something goes wrong because there is no way to fix or solder boards. The technology and the software is getting so advanced fewer people can actually work on them which is worrying.} [Week 3]
\end{quote}

Some participants found the balance between convenience and privacy an interesting shared discussion with their friends. Participant 19 was ambivalent about this:
\begin{quote}
\textit{I also discussed AI with my friends as we were discussing how personalized ads were becoming on Instagram and whether we liked it or not. We questioned whether trading privacy of being listened to for convenience of a relevant ad was something we cared about. We found that most people feel as though their privacy is already compromised so we might as well get relevant ads and Instagram posts out of it.} [Week 1]
\end{quote}

\subsubsection{\textbf{Interacting with Content}}
Content creation refers to media created for audiences and often uploaded to a social media platform for others to watch and engage. Participant 16 highlights:
\begin{quote}
\textit{I came across an intriguing video in which a YouTuber used AI to colorize old black and white photos from the 1900s. The video not only demonstrated the technological prowess of modern algorithms, but it also brought history to life, making it more relatable and vivid.} [Week 2]
\end{quote}

However, Participant 16 highlights the different angles across content:
\begin{quote}
\textit{However, not every mention of AI was rosy. A random episode of the podcast I listened to discussed the ethical quandaries associated with AI-powered facial recognition systems, highlighting potential privacy invasions and misuse. This discussion took on a more cautionary tone, reminding us that, while AI has enormous potential, it is also a two-edged sword.} [Week 2]
\end{quote}

\subsubsection{\textbf{Gaming}}
Gaming were mentioned several times across the journal entries in describing how AI could show up in unexpected places. Participant 16 highlights this by mentioning: 
\begin{quote}
\textit{The incorporation of AI in a flagship game like "Fortnite" demonstrates the gaming industry's evolving landscape. From the perspective of a gamer, it's both thrilling and fascinating. We're not just playing against a pre-programmed system anymore; we're navigating an ever-changing virtual world that's learning and reacting.} [Week 3]
\end{quote}

\textbf{\textit{Other General AI Observations}}

In addition to the categories highlighted in Table I, participants described interactions with AI that led to other general observations but did not specifically mention a source or event. Participants reflected by synthesizing and summarizing the formal elements of their learning and their feelings and opinions toward what they had interacted with during the week. Participant 6 raised the discussion about how AI is described broadly: 
\begin{quote}
\textit{A lot of the negative feedback I get about AI is it being abused in a way that will replace work or students using it to cheat. However, I think it opens up the conversation to look at the current processes we have, like how we teach students or how we can use it to complement the work that we do. Instead of all the doomsday negative impressions you get from those who struggle with change.} [Week 1] 
\end{quote}

Many of the participants scrutinized the benefits and costs of AI systems. As Participant 6 described, these feelings of navigating perceptions about the tools:
\begin{quote}
\textit{I feel like the usage of AI in a responsible manner would provide people with a way to make processes much easier. However, it's important to understand that quality controls are necessary to ensure that we are using these tools properly. With AI being a hot topic right now, I feel that I will continue to see articles about the public's reservations and doubts on AI and its usage. I also understand that not everyone understands how it works and its limitations.} [Week 4]
\end{quote}

In exploring the feedback, students had different expectations, experiences, and desires with AI. For some, the convenience factor was the ultimate decider to their attitude toward AI. Participant 19 sums this up with:
\begin{quote}
\textit{Overall this week I have come to believe that AI adds a sense of convenience to my life that I am happy accepting. The convenience should serve as a jumping off point for further personal research and not copying [regarding using AI for coursework].} [Week 1]
\end{quote}

\begin{figure}[htbp]
\centerline{\includegraphics[width=0.5\textwidth]{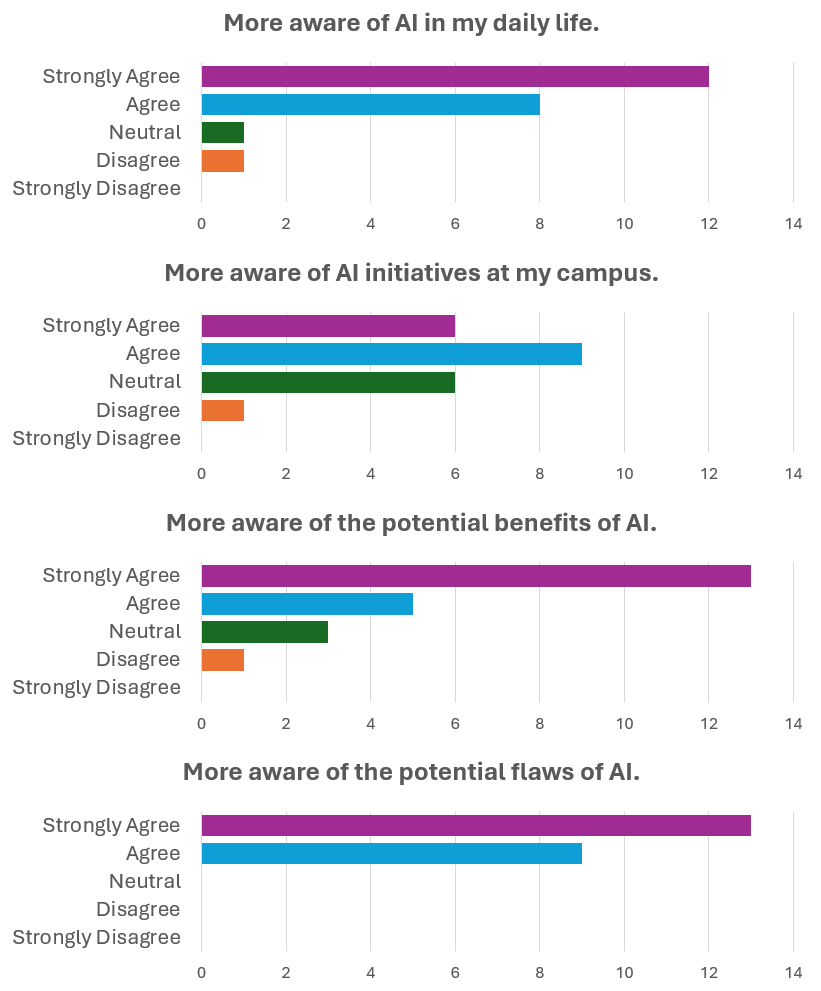}}
\caption{Student Responses to End-of-Study Survey}
\label{fig}
\end{figure}

\subsection{How did students perceive the process of reflecting on AI interactions? (RQ2)}

In addressing RQ2, we asked the participants to complete a short survey at the end of the reflective journal data collection. Figure 1 shows participant responses to the end-of-study survey. Overall, participants largely highlighted the reflection activities as helping them be more aware of all the aspects of the AI focus of this study. Participants highlighted being more generally aware of AI initiatives on their campus. 

\section{Limitations}

While this study provides valuable insights into expanding AI awareness through student's daily interactions with AI, it is important to acknowledge the limitations that may have influenced the results and interpretations. One of the primary limitations is the participants in this study were all pursuing a technology degree, which could have influenced their interactions with AI. This shared educational background likely provided them with a baseline understanding of technology and its applications. Consequently, their familiarity with technological concepts and systems might have shaped their responses and engagement with AI in ways that may not be representative of a broader population. 

Secondly, the emergent analysis methodology provides a flexible and adaptive way to explore qualitative data such as those collected through the reflective journal, but may lack the structured rigor of more predefined analytical frameworks, which is normal among studies of this style.

Finally, implementing this activity without the use of incentives may be challenging to do. In this study, incentives were provided for students' efforts to complete the reflective journals. The design of the study can be incorporated into a course through a graded assignment, but special attention would need to be given to the quality of responses.

\section{Future Work}
Thinking forward, this study highlights several important findings that can contribute to the direction of future work and instructional design. Firstly, interactions on campus through both formal and informal activities are an important mechanism for students to be exposed to conversations about AI and can supplement their learning in the classroom in a semi-structured way. However, the quality of engagements in this way is important.

Secondly, participants described interaction with media in content and gaming, highlighting areas that have largely not been explored in designing curriculum. It is important to note that these categories, even the less frequent interactions, were highly influential on many participants' perspectives toward AI. Some courses take advantage of popular culture themes, highlighting, for example, how students can learn about ethical speculation by having students participate in the \textit{Black Mirror Writers Room} \cite{fiesler2022blackmirror}.  Therefore, there is an opportunity to continue developing interesting activities aligning with topics students are excited about. However, this type of activity requires additional instructor effort and training \cite{Schleiss2024a, Schleiss2024b} and design of programs and courses that effectively integrate AI related knowledge.

\section{Conclusion}
Allowing students to reflect on their interactions with AI in a context beyond the classroom is important to ensure they build a holistic perspective. Using reflective journals as an unstructured learning activity is one way to have students do this reflection. Incorporating informal activities serves to anchor students' formal classroom learning experiences. Through reflection and integrating their individual perspectives, students demonstrate a more nuanced understanding of AI and connect it more broadly to other ideas. Moreover, informal learning approaches in AI provide a structured framework while accommodating flexibility, fostering an environment conducive to student-driven learning.

The participants in the study described nine distinct categories of interactions, spanning coursework, engagement with news and current events, utilization of software and applications, attendance at university events, involvement in social media platforms, work-related interactions, personal discussions with peers and family members, engagement with content, and participation in gaming activities. Through the reflection process, it became evident that students experienced heightened awareness regarding the prevalence of AI in their daily routines, along with a deeper understanding of its inherent strengths and limitations, as facilitated by the practice of journaling and reflecting on their interactions. Furthermore, students underscored the significance of the diary completion process, emphasizing its role in affording them valuable time for introspection and contemplation.

\section{Acknowledgments}
We would like to thank all the study participants for their invaluable time and effort, without which this study would not be possible. This work is partly supported by US NSF Awards 2319137, 1954556, and USDA/NIFA Award 2021-67021-35329. This work is also supported by a research grant through George Mason University's Mentoring for Anti-Racism and Inclusive Excellence (MARIE) program. Any opinions, findings, and conclusions or recommendations expressed in this material are those of the authors and do not necessarily reflect the views of the funding agencies.

\bibliographystyle{IEEEtran}
\bibliography{references}

\end{document}